\documentclass[preprint,12pt,sort&compress]{elsarticle}
\usepackage{amsmath,amssymb}
\usepackage{graphicx}
\usepackage{epstopdf}
\usepackage{color}
\usepackage{siunitx}

\journal{arXiv}

\begin{document}
\begin{frontmatter}

\title{On physical background of nerve pulse propagation: heat and energy}
\author{J\"uri Engelbrecht, Kert Tamm, Tanel Peets}
\address{Laboratory of Solid Mechanics, Department of Cybernetics, School of Science,\\ Tallinn University of Technology, Akadeemia tee 21, Tallinn 12618, Estonia, \\E-mails: je@ioc.ee, kert@ioc.ee, tanelp@ioc.ee}

\begin{abstract}
Recent studies have revealed the complex structure of nerve signals in axons. Besides the electrical signal, mechanical and thermal effects are also detected in many experimental studies. In this paper, the mathematical models of heat generation are analysed within the framework of a general model derived earlier by the authors. The main mechanisms of the heat generation are seemingly the Joule heating and endo- and exothermic reactions. The concept of internal variables permits to model the heat relaxation typical to these reactions. The general energy balance of the whole signal is analysed based on physical mechanisms responsible for emerging the components of a signal. Some open questions are listed for further studies. 
\end{abstract}
\begin{keyword}
Nerve signals \sep wave propagation \sep heat generation \sep energy transfer
	
\end{keyword}

\end{frontmatter}

\section{Introduction}

The basic component in signal propagation in nerve fibres is the action potential (AP). Since the pioneering studies of Hill \cite{Hill1932}, Hodgkin and Huxley \cite{Hodgkin1952}, FitzHugh \cite{FitzHugh1961} and many others, the electrophysiology of axons is well understood \cite{Clay2005,Debanne2011}. In addition, it is also well-known that the propagation of an AP is accompanied by mechanical and thermal effects (see overview by Watanabe \cite{Watanabe1986}). These effects are demonstrated by numerous experiments but theoretically there is no consensus about the possible coupling between the physical phenomena which generate a signal in axons as an ensemble of waves with several components. There are several hypotheses proposed which need not only the physiological explanations but also clear experimental proofs. In this situation, the best way is to return to basic physics,  from both theoretical and experimental viewpoints. One can agree with Heimburg \cite{Heimburg2020} that new instrumentation should be used for experiments, the time scale analysed, etc which could give new evidence on processes. He said: "Such experiments are absolutely necessary if one wants to understand nerves on a fundamental physical basis". Here, in this paper, theoretical ideas on modelling of processes in nerve fibres are analysed at the interface of electrophysiology, physics and mathematics. An attempt is made to describe the electrical, mechanical and thermal effects within the framework of a coupled mathematical model supported by physical considerations.
It relies on our recent paper \cite{Engelbrecht2020m} where the mechanisms of coupling were analysed. Here the focus is more on possible physical background and energy transfer between the components of an ensemble of waves. This ensemble is composed by:\\
 - the action potential (AP) and the corresponding ion current (J) or currents (J$_i$ 's);\\
 - the longitudinal wave (LW) in the biomembrane and the corresponding transverse deformation (TW);\\
 - the pressure wave (PW) in the axoplasm;\\
 - the temperature ($\Theta$) in the fibre.\\
In Section 2, the two paradigms for describing the processes in nerve fibres are described with a brief overview of mathematical models. The physical background of the processes is then analysed in Section 3. In Section 4, a brief description of possible modelling of temperature effects together with \emph{in silico} experiments is presented. Section 5 is devoted to the discussion and conclusions. 

\section{On modelling of processes in nerve fibres}
The classical Hodgkin-Huxley model \cite{Hodgkin1952} describes the action potential (AP) as an electrical signal which is generated after an electrical excitation and depends on the ionic mechanism (K and Na ions) of opening and closing the voltage-dependent channels in the axonal membrane (biomembrane). The propagation of an AP is then described by the cable theory (see, for example, \cite{Nelson2004,Bressloff2014}). No other effects are taken into account but the typical AP profile has been measured by many experiments and is a basic element in contemporary axon physiology \cite{Clay2005,Debanne2011}. If the K and Ca ions govern the ionic mechanism then the Morris-Lecar model \cite{Morris1981} is used with different from the Hodgkin-Huxley expressions for the phenomenological variables. The electrophysiological mechanism of generating an AP is supported by studies of electric synapses which cause voltage changes in the presynaptic cell transferred to postsynaptic cells. Here we leave aside how electrical synapses co-exist with chemical synapses.

This Hodgkin-Huxley model is sometimes called the Hodgkin-Huxley paradigm which means following the idea that the AP is the basic element in signal propagation. It is suggested that instead of this paradigm, a different basis may be used.

Namely, the Heimburg-Jackson model \cite{Heimburg2005} is based on the assumption that the basic signal is a mechanical wave (LW) in the membrane cylinder. The corresponding governing equation permits a soliton-type solution that is why this model is shortly called "the soliton model". According to authors explanations \cite{Heimburg2005}, such a model includes beside the longitudinal pulse in a membrane also the transverse swelling of the membrane and accompanying electrical and thermal effects. The generation of accompanying effects is explained theoretically but there are no governing equations describing the generation and propagation of these effects and only some ideas how a mechanical pulse in a biomembrane can be generated, are discussed \cite{HJ2007}. Another mechanically activated model was proposed by Rvachev \cite{Rvachev2010}. This model suggests that the propagation of the AP is driven by the pressure wave (PW) in axoplasm, which mechanically activates Na$^{+}$ ion channels resulting in a voltage spike. The velocity of the voltage spike is determined dependent on the fibre diameter. There are several studies supporting  "the soliton model" \cite{Vargas2011,Contreras2013,Barz2013} and discussions comparing the Hodgkin-Huxley and "soliton" models \cite{Appali2010,Drukarch2018}.

Engelbrecht et al \cite{EngelbrechtMEDHYP,Engelbrecht2018e} have proposed to return to basics and start from the physical considerations. This means that all dynamical processes should be governed by wave-type equations and all thermal processes should be based on the ideology of the Fourier law. Such an approach is widely used in continuum mechanics and in other words, means that the whole process of generation of a signal in an axon is characterised by single processes which are coupled into a whole. Metaphorically speaking, the single "bricks" (AP, J, LW, TW, PW and $\Theta$) are collected together using the coupling forces as bonds. Every "brick" has a clear and consistent physical basis. If the coupled system needs more complicated bonds between the "brick" or these bonds affect significantly the properties of "bricks" then it is easy to modify all the single constructions.

Following these ideas, Engelbrecht et al \cite{EngelbrechtMEDHYP,Engelbrecht2018e} start from the Hodgkin-Huxley paradigm which states that the basic component of signalling in nerve fibres is the AP. The main hypothesis for building the coupling forces is that other effects are generated by changes in the AP and ionic currents while there could also be coupling (feedback) between all the components which also depends on changes of variables. In mathematical terms, the changes are described by space or time derivatives of variables. A pulse-type profile of a variable means a bi-polar shape of the derivative which is energetically balanced. The possible physical interpretation of such derivatives are described in \cite{Engelbrecht2020m}. It is important to notice that such an approach does not require a specific model for generating the AP. The basic requirement for modelling the AP is to get a correct shape of it together with the corresponding ion current(s). Engelbrecht et al \cite{Engelbrecht2018e} have used the FitzHugh-Nagumo (FHN) model \cite{Nagumo1962} with one generalised ionic current in their numerical simulations but showed also how to use the Hodgkin-Huxley (HH) model with two (sodium and potassium) ionic currents \cite{Engelbrecht2020a}.  In principle, it is possible to use an experimentally measured AP only then the question is about the ion currents. Leaving open the source of generating the AP in modelling, the question of adiabaticity of one or another model of the AP is in principle not so relevant.  The main emphasis in modelling described briefly above, is to build up suitable governing equations for accompanying effects and plausible coupling forces (see \cite{EngelbrechtMEDHYP,Engelbrecht2020m}). The numerical simulations have demonstrated good qualitative similarity to various experimental results  \cite{Engelbrecht2018e,Engelbrecht2018,Engelbrecht2019b}. 

El Hady and Machta \cite{ElHady2015} follow also the Hodgkin-Huxley paradigm and take (without calculations) the AP as a Gaussian pulse for a driving force to mechanical and thermal effects. They use an assumption that the biomembrane stores the potential energy of the whole system and the axoplasm - the kinetic energy. Chen et al \cite{Chen2019} have used the mechanism of flexoelectricity for explaining the transverse deformation of the biomembrane as a result of the propagating AP. 

\section{On physical background}
It is a general understanding in all the models that wave motion is the basic process in signal propagation although the mathematical models are significantly modified to reflect the physics. The discussions on the thermal part of the signal, ie heat production and temperature change are going on.  Whatever the mechanism of generating thermal effects is, following basic physical considerations, all thermal changes should this way or another be related to the Fourier law. This is characteristic to all the bioheat processes \cite{Pennes1948}. In principle, the whole process nerve pulse propagation should be adiabatic \cite{Kaufmann1989,Appali2010,Heimburg2020}. However, all real physical processes are dissipative and according to Margineanu and Schoffeniels \cite{Margineanu1977}, "part of the free energy is degraded into heat". The role of dissipative processes is also stressed by Kaufmann \cite{Kaufmann1989}. It means that energetics of changes occurring in fibre during the signal propagation should be analysed in terms of all the components of an ensemble.

The problem of coupling of components in an ensemble must be understood first in physical terms and then in a governing model, in mathematical terms. The main hypothesis proposed by Engelbrecht et al \cite{EngelbrechtMEDHYP} is: the mechanical waves in axoplasm and surrounding biomembrane are generated due to changes in electrical signals (the AP or ion currents). In case of thermal effects, the same hypothesis is used \cite{Engelbrecht2020a} but needs some modifications due to the possible existence of several mechanisms (see below). Such a hypothesis, although not directly, is mentioned by Margineanu and Schoffenniels \cite{Margineanu1977} as "changes in the electrical field" and certainly could be traced back to Emil Du Bois-Reymond, who mentioned the "variation" acting as a stimulus \cite{Hall1999}. From a mathematical viewpoint, a change of a variable means its spatial or temporal derivative. If the profile of a variable has a pulse-type shape then the derivative is bi-polar and such derivatives are used as the first approximations for coupling forces \cite{EngelbrechtMEDHYP,Engelbrecht2018e,Engelbrecht2018}. This idea is following the solution of a wave equation with a driving force which is energetically balanced. 

\begin{figure}[h]
\includegraphics[width=0.20\textwidth]{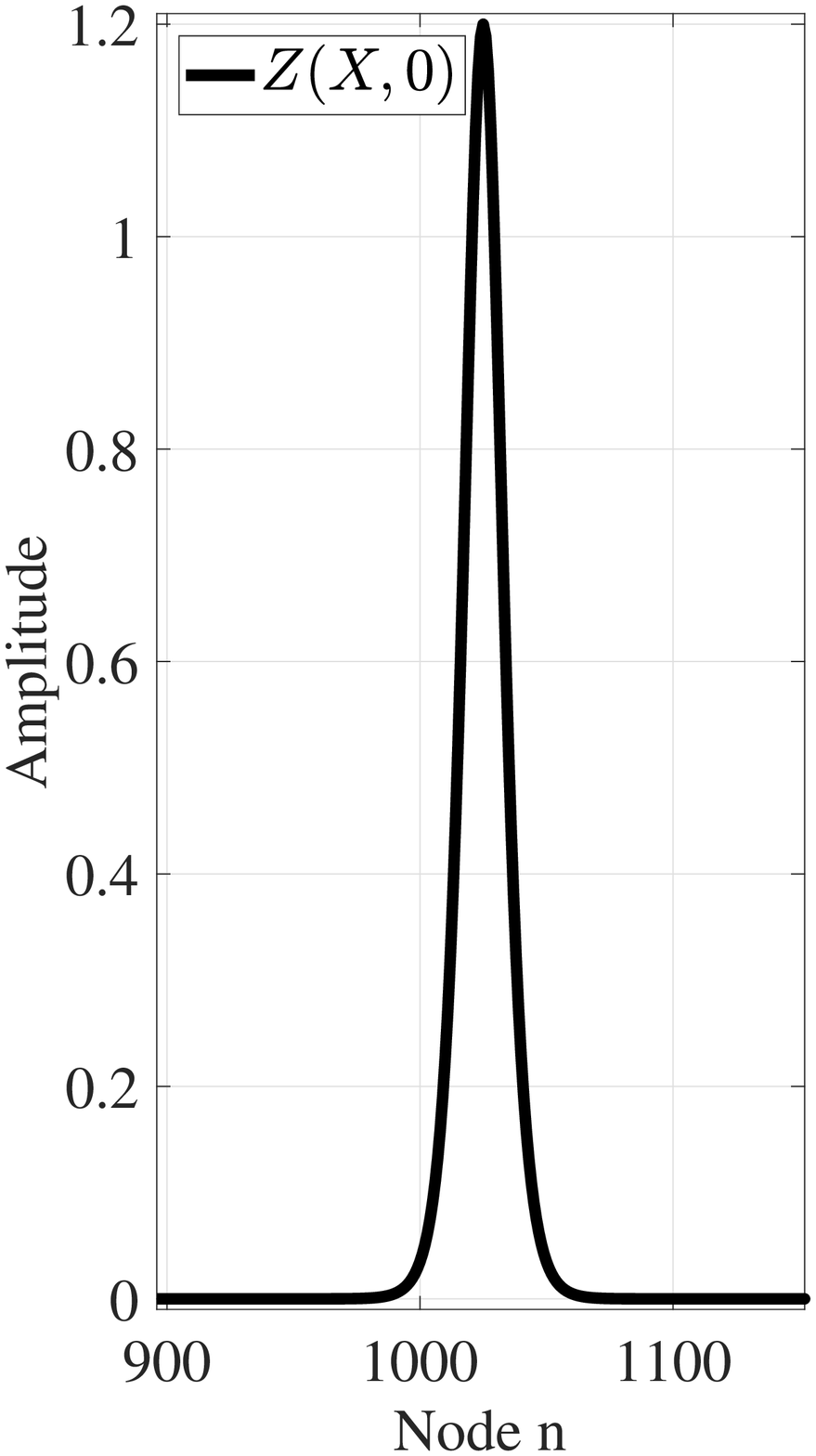}
\includegraphics[width=0.24\textwidth]{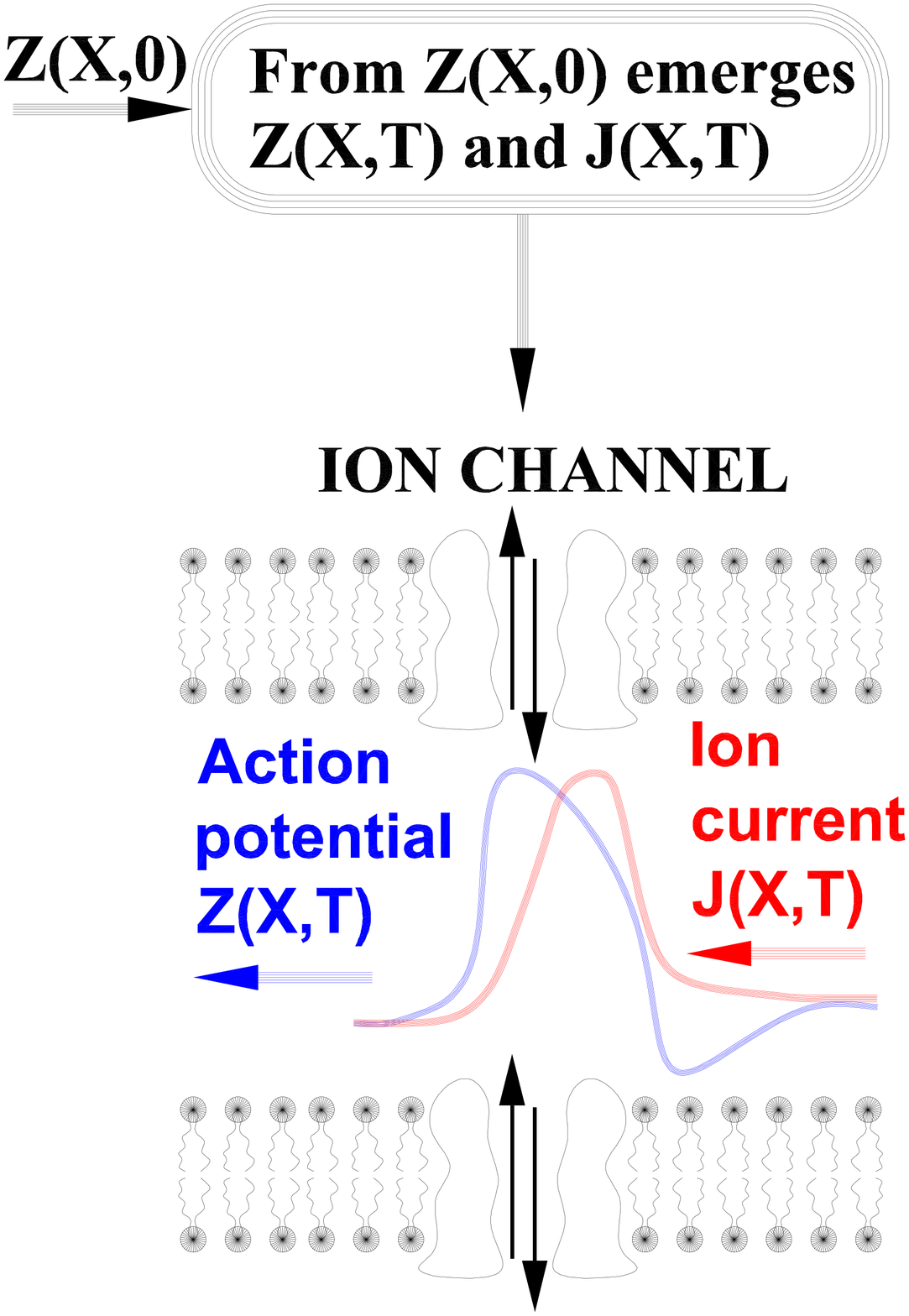}
\includegraphics[width=0.49\textwidth]{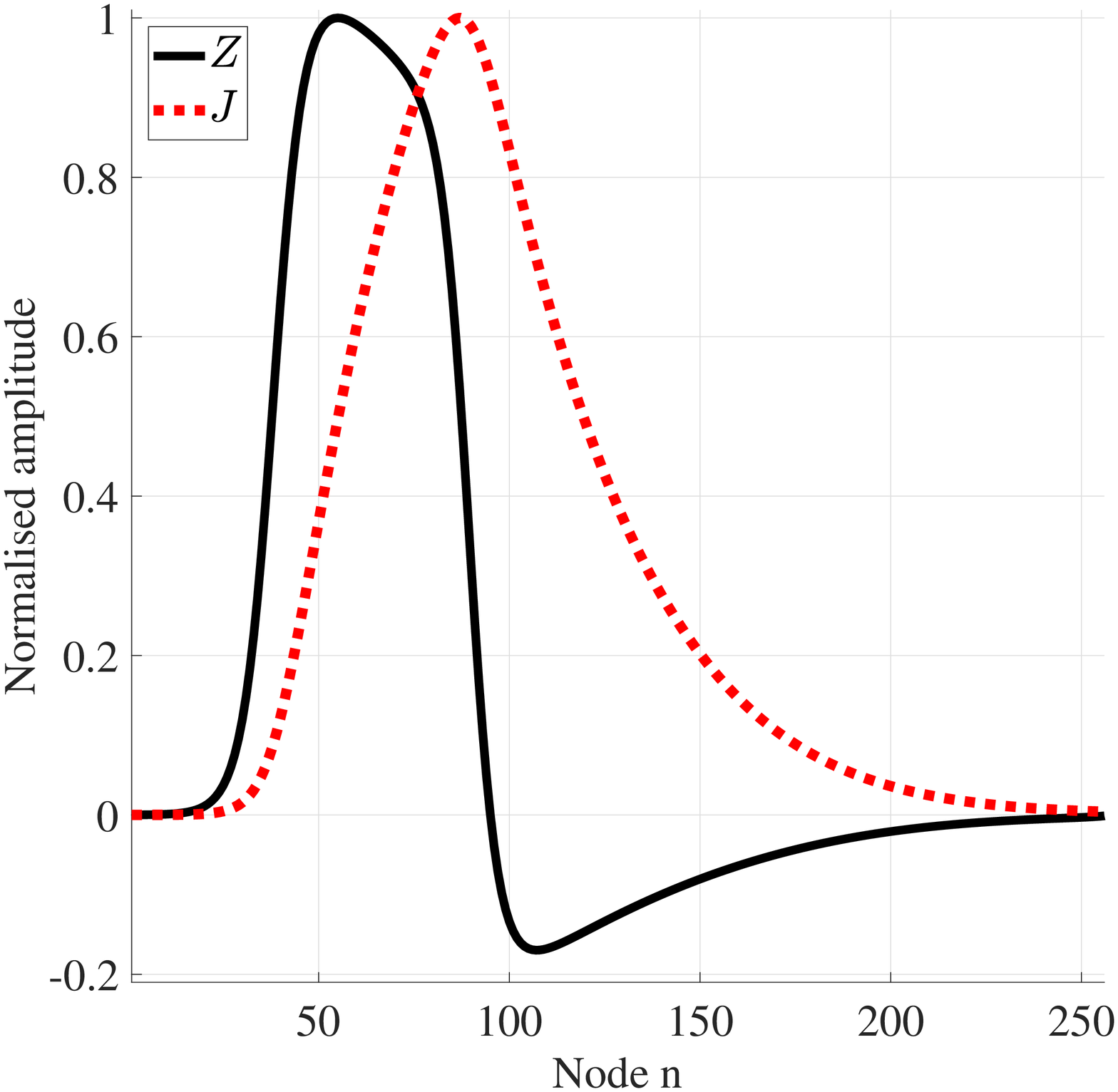}
\caption{The initial condition (left), the solution scheme (middle) and the typical numerical solution of the FHN model (right). The solid black line denotes the action potential $Z$ while red dotted line represents the ion current $J$.}
\label{fig1}
\end{figure}

\begin{figure}[!h]
\centering
\includegraphics[width=0.70\textwidth]{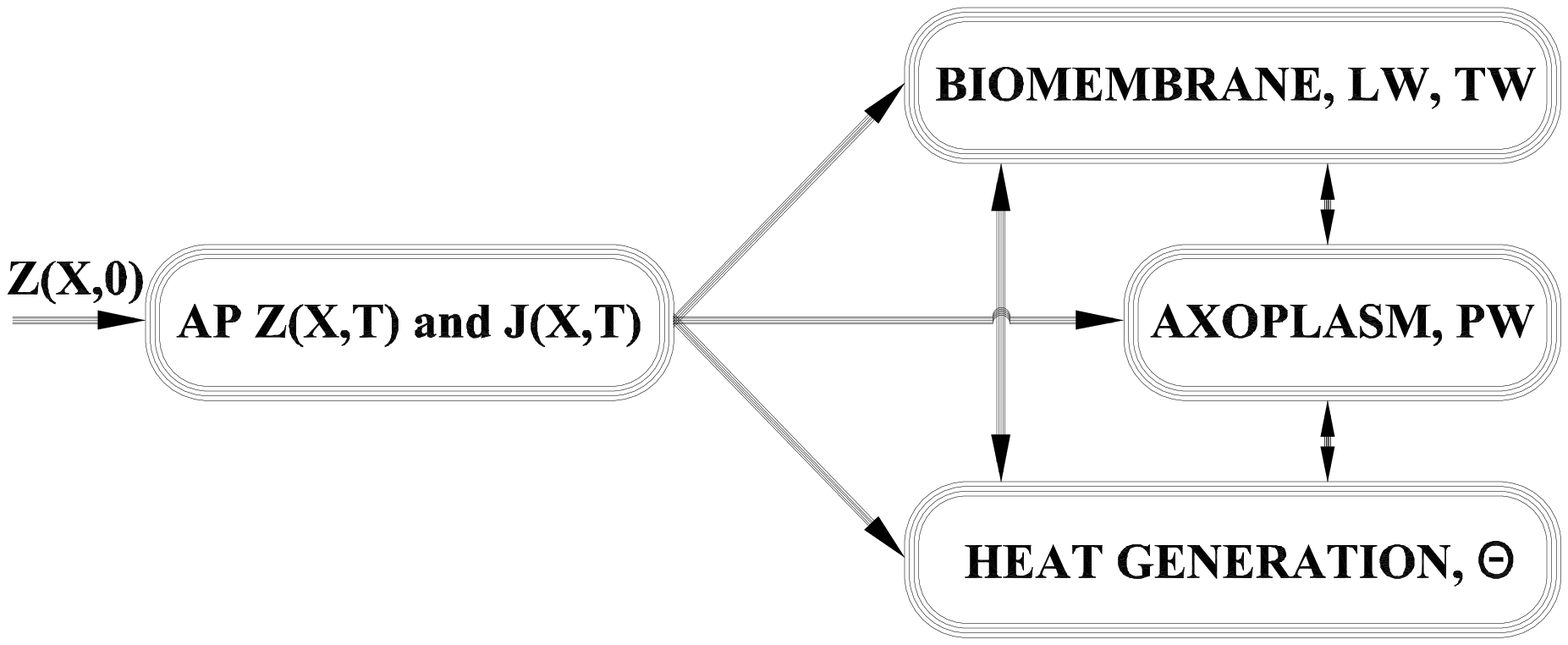}\\
\includegraphics[width=0.90\textwidth]{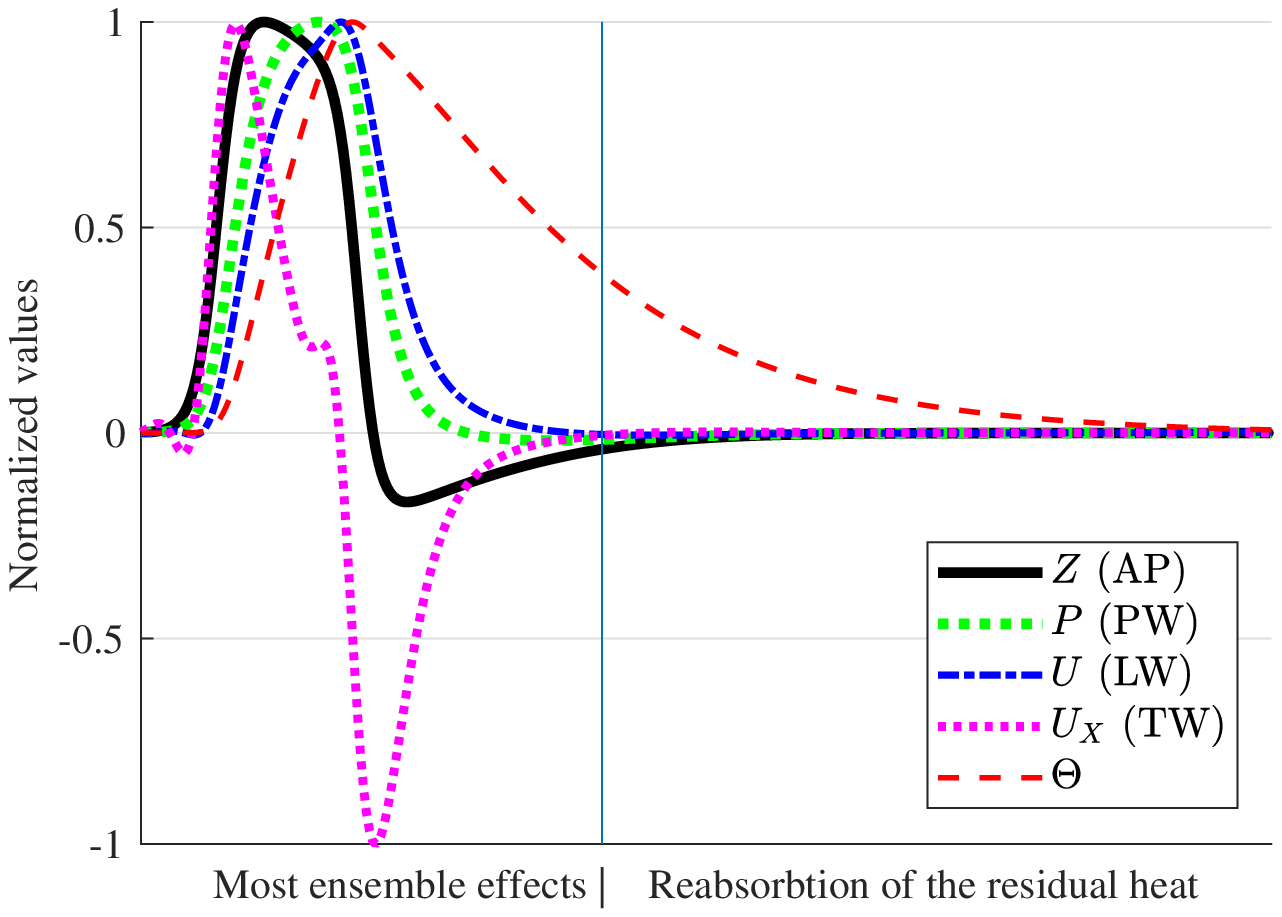}
\caption{The AP with other effects. Top: the block-scheme of the wave ensemble and the interactions. Bottom: the typical dimensionless numerical solution of the wave ensemble.}
\label{fig2}
\end{figure}

The AP is generated by a synaptic input in the initial segment of an axon at a certain distance from the cell body \cite{Shu2007,Bean2007}. This distance in mammalian neurons could be about 30-50 $\mathrm{\mu}$m and due to various synaptic behaviours, the diversity of formed spikes has been measured \cite{Bean2007}. In experiments, an electric input is used (see, for example, \cite{Hodgkin1945,Tasaki1988,Yang2018}) to generate a propagating AP. The process of forming an AP (amplitude $Z$) is shown in Fig.~\ref{fig1} where the ion current $J$ calculated by the FHN model is similar to the one presented in \cite{Engelbrecht2018c}. 
The propagating AP together with the ion current (or currents if the HH model is used) is the source for other effects. This means forming an ensemble of waves which as a result of numerical simulation \cite{Engelbrecht2019b} is presented in Fig.~\ref{fig2}. It is seen that the LW, TW, PW are in phase with the AP like the numerous experiments have shown (see analysis by Engelbrecht et al, \cite{Engelbrecht2019b,Tamm2019,Engelbrecht2020m}). The behaviour of heat generation and the corresponding temperature profile is however different and there is still no consensus about the basic mechanisms of heat generation. 
A brief overview of corresponding experiments and suggested ideas leads to the conclusion that there might be several mechanisms responsible for this phenomenon. The early experimental studies of Hill \cite{Hill1932}, Abbott et al \cite{Abbott1958}, Howarth et al \cite{Howarth1968} have all demonstrated that the heat generation by an AP can be characterised by the fast period (in phase with the active phase of an AP) and the slow period which is related to the absorption of heat. The fast period is called the initial heat \cite{Hill1932,Ritchie1985} or positive heat \cite{Howarth1968} and the slow period - negative heat \cite{Howarth1968}. The slow period may last about 240 msec for non-myelinated rabbit vagus nerves \cite{Howarth1968}. It is however estimated \cite{Downing1926} that about 90\% of the total heat is liberated after the stimulus ie the AP is over. The general understanding is that the heat generated during the fast period may be the result of the Joule heating \cite{Howarth1968,Nogueira1983a} or capacitative processes \cite{Howarth1979,Ritchie1985,Tamm2019} and the process of reabsorption may be due to endothermic chemical reactions \cite{Abbott1958}.

\section{Modelling of temperature effects}
There are several suggestions on how to calculate the generated heat or directly the temperature change. 
Tasaki and Byrne \cite{Tasaki1992} proposed to use a polyvinylidene fluoride film as a detector to measure the heat generated in the bullfrog sciatic nerve. They calculated the temperature from the following equation (in original notations):
\begin{equation}
\label{eq1}
C \left(\frac{\mathrm{d} V}{\mathrm{d} t}\right) + \frac{V}{R} = sp \left(\frac{\mathrm{d} T}{\mathrm{d} t} \right)
\end{equation}
where $V$ is the output voltage, $T$ is the temperature, $C$ and $R$ are the feedback capacity and resistance of the amplifier, $s$ is the area and $p$ - the pyroelectric constant of the film. Depending on product $RC$, (long or short in ms) either one or another term at the LHS is taken into account. According to this equation, the temperature is related to voltage.

El Hady and Machta \cite{ElHady2015} link the heat to the AP  over mechanical effects summing the longitudinal and transverse effects or in general, relating heat $Q$ to $h(z)+l(z)$. Here $h(z)$ is the relative (transverse) height (displacement) of the membrane and $l(z)$ is the lateral stretch, while $z$ is the longitudinal coordinate.  Their calculations show that the temperature profile is an asymmetric pulse with an overshoot. So actually following El Hady and Machta \cite{ElHady2015}, the AP generates mechanical effects which in their turn generate the heat and the temperature.

According to Heimburg \cite{Heimburg2020}, the temperature increment depends on the adiabatic compression close to melting transition in biomembranes (in original notations):
\begin{equation}
\label{eq2}
\mathrm{d}T=-\frac{T}{c_{A}}\left(\frac{\alpha_{\Pi}}{\kappa_{T}^{A}}\right)\mathrm{d}A
\end{equation}
where $T$ is the temperature, $A$ is the area, $c_{A}$ is heat capacity, $\alpha_{\Pi}$ is the thermal area expansion coefficient and $\kappa_{T}^{A}$ is the isothermal area compressibility. Later it is linked the latent heat of the membrane transition. So here temperature is linked to the changes and properties of the membrane, not to voltage or reactions.

We suggest modelling the temperature change using the classical heat equation with a driving force \cite{Tamm2019,Engelbrecht2020m,Engelbrecht2020a}. This idea permits to model several possible mechanisms responsible for thermal effects. Following this model, the temperature is governed by the classical heat equation with an external force F:
\begin{equation} 
\label{eq3}
\Theta_{T} = \alpha \Theta_{XX} + \mathrm{F}(Z,J,U,P),
\end{equation}
\begin{equation} 
\label{eq32}
\mathrm{F} = \tau_{1} Z^{2} + \tau_2 \left( P_T + \varphi_2(P) \right) + \tau_3 \left( U_T + \varphi_3(U) \right)-\tau_4\Omega,
\end{equation}
\begin{equation} 
\label{eq33}
\Omega_T = \varphi_4(J) - \frac{\Omega-\Omega_0}{\tau_\Omega},
\end{equation}
\begin{equation} 
\label{eq34}
\varphi_2(P) = \lambda_2 \int P_T \mathrm{d}T, \, 
\varphi_3(U) = \lambda_3 \int U_T \mathrm{d}T, \,
\varphi_4(J) = \lambda_4 \int J \mathrm{d}T,
\end{equation}
where $\Theta$ is the temperature, $\alpha$ is the thermal conductivity coefficient and F is the external force depending on amplitude $Z$ of the AP, amplitude $P$ of the pressure wave PW and amplitude $U$ of the longitudinal wave LW. Indices $X, T$ here and further denote the differentiation with respect to space and time, respectively while index $1$ is used for quantities associated with the AP, index $2$ is used for quantities associated with the PW, index $3$ is used for quantities associated with the LW and index $4$ for quantities associated with the internal variable(s) describing thermal processes happening in the different time-scale than the driving signal (which is the AP here). In Eqs \eqref{eq32} and \eqref{eq34} the $\tau_i$ and $\lambda_i$ are coefficients while $\varphi_i$ are integral-type terms accounting thermal influence from the irreversible thermal processes (temperature increase from the energy lost to the dissipation from the mechanical waves and endo-- or exothermal chemical reactions for the $\varphi_4$). Equations \eqref{eq3} and \eqref{eq32} are able to model several effects. First, the Joule heating/thermal influence of the capacitor energy is taken into account by including the term $\tau_{1} Z^{2}$ while the term involving parameter $\tau_2$ accounts for the thermal influence from the pressure wave in axoplasm and the term involving parameter $\tau_3$ accounts for the thermal influence from the mechanical wave in a biomembrane. Second, for the accounting of the possible endothermic reactions \cite{Abbott1958} an internal variable $\Omega$ is introduced \cite{Engelbrecht2020a} following Maugin and Muschik \cite{Maugin1994a}. This internal variable $\Omega$ is governed by a kinetic equation \eqref{eq33} where $\tau_\Omega$ is the relaxation time, $\Omega_0$ denotes the equilibrium level and $\varphi_4$ characterises the thermal influence of a `concentration-like quantity' decaying exponentially in time \cite{Tamm2019}.
Such a presentation of coupling force F permits to model slow relaxation of temperature towards equilibrium \cite{Abbott1958,Engelbrecht2020a}. If the HH model is used with two ion currents, then two internal variables must be introduced \cite{Engelbrecht2020a}. In this case, the influences of exo-and endothermic effects can be clearly separated. 

\section{Discussion}
The propagation of a nerve impulse is a complex process spiced with nonlinearities. In order to understand the neural responses of the brain to various stimuli, one should also understand the physical mechanisms of how signals propagate in the huge network of neurons. The basic element of this network is a single fibre where not only an electrical signal propagates but one should pay attention to an ensemble of waves. This means the analysis of coupled effects must be based on electromechanophysiological interactions and thermal effects resulting in the physical reality which is "far richer than that assumed in a pure electrical picture" \cite{Heimburg2020}. 

One possible mathematical model is presented by Engelbrecht et al \cite{EngelbrechtMEDHYP} based on the coupling of single effects into a whole. The idea is to start from basic physics and corresponding experiments for deriving the governing equations and coupling forces (see assumptions by Engelbrecht et al, \cite{EngelbrechtMEDHYP,Engelbrecht2019b,Engelbrecht2020m}). Basic physics means that whatever the modifications of governing equations, the starting point is related to wave equations (dynamics) and Fourier's law (temperature). Based on the ideas of continuum mechanics, the concept of internal variables is used for describing the exo- and endothermic reactions \cite{Engelbrecht2020m,Engelbrecht2020a}. The numerical simulation has demonstrated a good match with experimental data. Beside coupling forces, one should pay attention also to general energetical balance of the whole process. The general understanding is that the whole process of signal propagation in nerve fibres is adiabatic and reversible. However, the experiments have demonstrated that there is residual heat after the passage of an AP. This is demonstrated for mammalian (rabbit) nerves \cite{Howarth1968} and fish (pike) nerves \cite{Howarth1975}, in both cases about 10\% of positive initial heat. Heimburg \cite{Heimburg2020} stated that "from the perspective of an energy, the heat change is in fact larger than the electrical effect, even if temperature changes are small".

\begin{figure}[h]
\centering
\includegraphics[width=0.70\textwidth]{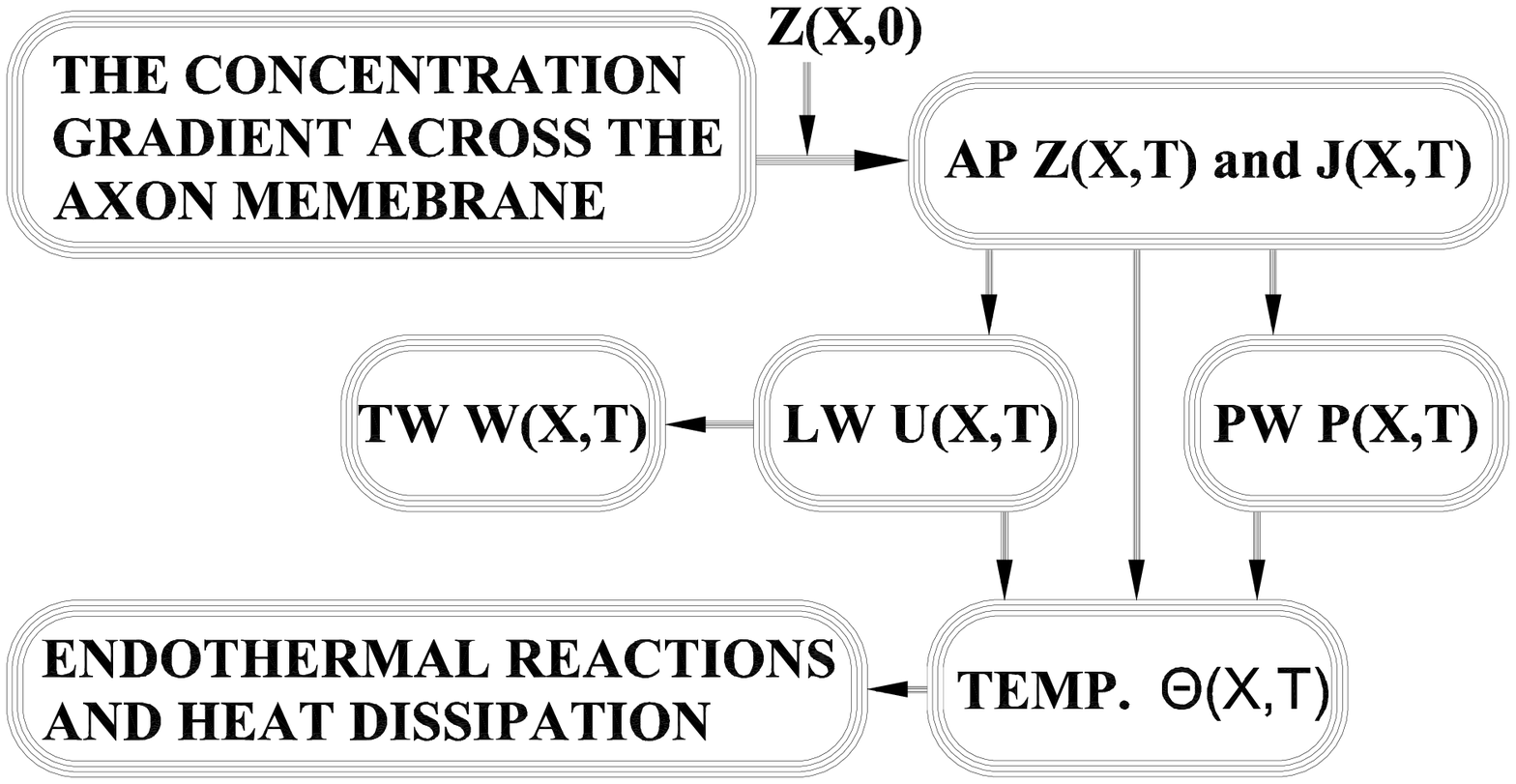}
\caption{The energy flow of the proposed model framework.}
\label{fig3}
\end{figure}

The energy flow within the proposed model framework is schematically shown in Fig.~\ref{fig3}.
The source of energy for all the modelled processes is the concentration gradient across the axon membrane. The initial excitation $Z(X,0)$ with an amplitude above the threshold value has some energy as well but is only a trigger leading to an emergence of a propagating action potential and associated ion current(s) which, in turn, transfers energy to the mechanical waves (LW in biomembrane and a PW in axoplasm). The TW is not independent but in the present framework is derived from the energy of the LW. The mechanical waves lose some of their energy as heat through friction/viscosity associated dissipative terms and some of the energy of the AP is transferred directly to the heat as well through Joule heating/capacitative processes. Then the heat energy is removed from the model through diffusive processes (which could be assumed to be too slow to transfer a significant amount of heat out of the model within the time-scales under consideration) and through abstracted "endothermic reactions" which can consume heat locally. It must be noted that the processes which maintain and restore the concentration gradient across the axon membrane are not included in the model framework but it is, in essence, a naive assumption that this gradient is maintained and is restored fully after certain (relaxation) time when the AP model allows one to start the next propagating AP pulse. Expanding the proposed model framework by a dynamic description of the processes restoring the concentration gradient across the biomembrane is a possibility for scenarios where the existence of the noted gradient can not be assumed to be always fulfilled, for example, if repeating high-frequency excitation is modelled.

The challenge is to understand the general energy $E$ balance:
\begin{equation}
\label{eq9}
E = E_\mathrm{AP} + E_\mathrm{PW} + E_\mathrm{LW} + E_{\Theta},
\end{equation}
where indices denote the corresponding waves and temperature. Note that energy of ion currents \cite{Margineanu1977} is actually hidden in the $E_{\mathrm{AP}}$ because actually they play role of internal variables in supporting the AP. The energy  $E_{\mathrm{AP}}$ of an AP is \cite{Ritchie1985,Heimburg2008} 
\begin{equation}
\label{eq10}
E_\mathrm{AP} = \frac{1}{2}C Z^2
\end{equation}
where $C$ is the membrane capacitance and $Z$ is the amplitude of the AP. The same expression \eqref{eq10} can describe also the lipid bi-layer membrane tension change \cite{Terakawa1985} as a result of the voltage change. In general, energy should be related to motion, ie kinetic energy (see also Barz et al, \cite{Barz2013}). For  wave equations it means the  dependence on the square of the amplitude $A$ (ie on $A^2$ ) (see Margineanu and Schoffeniels \cite{Margineanu1977} for energy of ion currents). Heimburg and Jackson \cite{Heimburg2005} have explicitly proposed that for the LW according to their model, the energy is
\begin{equation}
\label{eq11}
E_\mathrm{LW} = \frac{c_{0}^{2}}{\rho_{0}^{A}}U^2 + \frac{p}{3} \rho_{0}^{A} U^3 + \frac{q}{6} \rho_{0}^{A} U^4,
\end{equation}
where $c_{0}$ is the velocity, $\rho_{0}^{A}$ is the density, $U=\Delta \rho_{0}^{A}$ is the amplitude and $p, q$ are the nonlinear parameters (see also Mueller and Tyler, \cite{Mueller2014}). Note that nonlinearity affects also the energy. The more general analysis of energy in biomembranes \cite{Deseri2008} in terms of continuum mechanics has divided the surface Helmholtz energy to local and non-local components. The local components depend on deformation and the non-local ones - on curvatures of the biomembrane. El Hady and Machta \cite{ElHady2015} proposed that the energy of the biomembrane is related mostly to potential energy $U_{\mathrm{LW+TW}}$  and the energy of the axoplasm - to kinetic energy $T_{\mathrm{PW}}$.

It is also known that the energy in an electromagnetic wave is proportional to the square of its peak electric field. The dissipation needed for describing the real processes \cite{Kaufmann1989} is included also to the modified wave equations governing the longitudinal waves in biomembranes and axoplasm. As said by Margineanu and Schoffeniels \cite{Margineanu1977} for ionic currents in the HH model, energy dissipation is degraded into heat. The natural phenomena as described above in Eqs \eqref{eq3} to \eqref{eq34} are dissipative by nature. So, from the viewpoint of energy balance, the coupled model derived by Engelbrecht et al \cite{Engelbrecht2020m} is able to redistribute energy between its components although we presently do not know the transduction of energy in mathematical terms. 
Within the coupled model it is obvious that the coupling is characterised by the energy transfer
between the components of the wave ensemble. For example, in the case of the LW, expression \eqref{eq11}
describes the conservative situation for an LW only. The total balance for the mechanical waves in the biomembrane (the LW and TW) during the propagation of the coupled signal could be described by
\begin{equation}
\label{eqLWenergy}
E_{\mathrm{LW+TW}}^{\mathrm{total}} = E_{\mathrm{LW}} + E_{\mathrm{LW}}^{\mathrm{coupl.}} - E_{\mathrm{LW}}^{\mathrm{dissip.}},
\end{equation}
where $E_{\mathrm{LW}}^{\mathrm{coupl.}}$ denotes energy inflow (through coupling force $F_3$) and $E_{\mathrm{LW}}^{\mathrm{dissip.}}$ denotes the energy loss from dissipation and through possible coupling with other components (for example, if energy exchange between PW and LW is accommodated). The similar arguing concerns the PW and $\Theta$.

One must agree with Heimburg \cite{Heimburg2020} that new experiments are needed because there are many unanswered questions.
First, more detailed analysis is needed to explain the energy transfer between the components of the ensemble of waves in a nerve fibre. Second, the experiments should give an answer on main mechanisms responsible for heat generation, although presently the Joule heating and the influence of exo- and endothermic reactions seem to be the most important mechanisms. Third, the influence of residual heat measured in experiments must be understood in more details. For example, there is a question how the slow period of reabsorption of heat affects the generation of a next AP? Fourth, the processes in biomembranes should be analysed from a viewpoint of the emergence of possible solitons like it is explained in soliton theory \cite{Ablowitz2011} including also the role of a soliton in forming an AP with an overshoot. 

One should stress that the modelling described above relies certainly only on basic physical effects. The reality is much more complicated because physiological details of nerves, as demonstrated by numerous experiments can strongly influence the formation of the main carrier of information, the AP \cite{Bean2007,Debanne2011}. Working at the interface of physics and physiology \cite{Berg2014} and using mathematical modelling \cite{Gavaghan2006}, could enrich the understanding of nerve pulse dynamics. The new experimental studies like by Lee et al \cite{Lee2019} on small-scale membrane displacements could certainly give more evidence about the processes.  

\section*{Acknowledgments} This research was supported by the Estonian Research Council (IUT 33-24).


\end{document}